\documentclass[aps,prd,nofootinbib,twocolumn,superscriptaddress]{revtex4}

\usepackage[utf8]{inputenc}
\usepackage{amssymb,amsmath,slashed}
\usepackage{graphicx,comment,multirow,subfigure,tabularx}
\usepackage{placeins}
\usepackage{verbatim}
\usepackage{makecell}
\usepackage{hyperref}
\usepackage{booktabs}
\usepackage{enumerate} 
\usepackage{color}

\usepackage[titletoc]{appendix}

\newcommand{\nicpb}{National Institute of Chemical Physics and Biophysics, R\"avala 10, 10143 Tallinn, Estonia}
\newcommand{\ut}{University of Tartu, Institute of Physics, Ravila 14c, 50411 Tartu, Estonia.}
\newcommand{\triestea}{Dipartimento di Fisica, Sezione Teorica, Universit\`a degli Studi di Trieste\\Strada Costiera 11, I-34151 Trieste, Italy}
\newcommand{\triesteb}{Dipartimento di Fisica, Universit\`a degli Studi di Trieste, 
Via Valerio 2,  I-34127 Trieste, Italy.}
\newcommand{\infn}{INFN, Sezione di Trieste, Via Valerio 2,  I-34127 Trieste, Italy.}

\newcommand{\milli}{\epsilon}
\newcommand{\Mmilli}{m_{\epsilon}}

\newcommand{\td}{\mathrm{d}}

\newcommand{\bea}{\begin{eqnarray}}
\newcommand{\eea}{\end{eqnarray}}

\newcommand{\gsim}{\lower.7ex\hbox{$\;\stackrel{\textstyle>}{\sim}\;$}}
\newcommand{\lsim}{\lower.7ex\hbox{$\;\stackrel{\textstyle<}{\sim}\;$}}

\newcommand{\atanh}{\mathrm{th}^{-1}}

\newcommand{\s}{{\scriptscriptstyle{S}}}

\begin{document}

\title{Polarization observables for millicharged particles in photon collisions}

\author{Emidio Gabrielli}
\affiliation{\triestea}
\affiliation{\infn}
\affiliation{\nicpb}

\author{Luca Marzola}
\affiliation{\nicpb}
\affiliation{\ut}

\author{Edoardo Milotti}
\affiliation{\triesteb}
\affiliation{\infn}

\author{Hardi Veerm\"ae}
\affiliation{\nicpb}

\date{\today}


\begin{abstract}
Particles in a hidden sector can potentially acquire a small electric charge through their interaction with the Standard Model and can consequently be observed as millicharged particles. We systematically compute the production of millicharged scalar, fermion, and vector boson particles in collisions of polarized photons. The presented calculation is model independent and is based purely on the assumptions of electromagnetic gauge invariance and unitarity. Polarization observables are evaluated and analyzed for each spin case. We show that the photon polarization asymmetries are a useful tool for discriminating between the spins of the produced millicharged particles. Phenomenological implications for searches of millicharged particles in dedicated photon-photon collision experiments are also discussed.
\end{abstract}
 
\maketitle

\section{Introduction}

Although we observe electric charge quantization in nature, this property is not a requirement of the Standard Model (SM) \cite{Foot:1990mn}. In fact, whereas new physics models based on grand unified theories \cite{Georgi:1974sy} or proposing magnetic monopoles \cite{Dirac:1931kp}  enforce the charge quantization, other theories of physics beyond the SM predict the existence of particles with an electric charge $\milli \ll 1$.\footnote{Here and in the rest of the paper, the electric charge is measured in units of the elementary charge $e$.} The particles characterized by these possibly effective nonquantized charges are commonly dubbed millicharged particles (MCP) \cite{Ignatiev:1978xj,Holdom:1985ag,Abel:2003ue,Batell:2005wa}.

The most natural framework that yields MCP presents two or more $U(1)$ gauge groups, coupled to different matter sectors, whose fields possess nondiagonal kinetic terms. These induce the mixing of the corresponding gauge bosons, and, as a result, the matter fields of one sector appear as MCP in the other. We remark that even in the absence of a tree-level kinetic mixing, a nondiagonal kinetic term is inevitably induced by radiative corrections \cite{Holdom:1985ag},
provided there are matter fields charged under both $U(1)$ gauge groups.

The above mechanism can give rise to spin-0 and spin-1/2 MCP, but generating a consistent and unitary theory for elementary spin-1 MCP requires a different construction. In fact, while the interactions of scalars and fermions with the photon can always be induced via the mentioned mixing, the spin-1 case requires the extension of the SM group to a larger non-Abelian gauge group \cite{Gabrielli:2015hua}, with the spin-1 MCP ($V^{\mu}$) arising from the vector boson multiplet of the latter. 

The case of spin-1 MCP also presents an intriguing feature: As a result of the interplay between gauge invariance and unitarity, the total cross section of $\gamma \gamma \to VV$ tends to a constant in the high energy limit, whereas the same quantity decreases as $s^{-1}$ in the cases of spin-0 and spin-1/2 MCP. Such a distinguishing characteristic of spin-1 interactions is manifest in the SM, where the tree-level total cross section for $\gamma\gamma \to W^{+} W^{-}$ approaches a constant of about 80~pb at high energies \cite{Pesic:1973fi,Ginzburg:1982bs,Katuya:1982ga}, while radiative corrections are typically of order 10\% \cite{Denner:1995jv}.
This property, resulting from a collinear effect of the $W$ propagator in the $t$-channel production mechanism, has a general validity and also holds for the analogous production of two spin-1 MCP \cite{Gabrielli:2015hua}. Indeed, the requirement of unitarity and gauge invariance completely fixes the interaction Lagrangian of spin-1 MCP $V_{\mu}$ with the photon field, which necessarily recovers the structure of the $W^{\pm}$ Lagrangian of the SM, implying a gyromagnetic factor $g_V=2$ for the former.

In principle, the different asymptotic behavior of the spin-1 MCP production cross sections in photon-photon collisions could then be used as a tool to disentangle the production of these particles from the spin-0 and spin-1/2 cases. However, an even better sensitivity to the spin of the produced MCP particles could be achieved by employing polarized photon beams. In this case, as we will demonstrate, suitable polarization observables yield sensitive tools to probe the spin of MCP.

The most stringent limits on MCP models come from cosmological and astrophysical observations that bound the ratio of the millicharge fraction $\milli$  to the MCP mass $\Mmilli$. These constraints are model dependent and mainly apply to models where millicharges arise as a consequence of kinetic mixing \cite{Vogel:2013raa}. For MCP of mass below the MeV scale, the most relevant constraints come from stellar evolution and cosmology. For instance, as the emission of MCP pairs with low mass could induce a severe energy loss in stars, stellar evolution constrains $\milli < {\cal O}(10^{-14})$
for $\Mmilli < {\cal O}(10 {\rm KeV})$. The requirement of successful big bang nucleosynthesis leads, instead, to  $\milli < {\cal O}(10^{-9})$ for $\Mmilli < {\cal O}(10 {\rm MeV})$ \cite{raffelt-book}. Besides, if MCP can be considered charged dark matter constituents, assumptions on the magnetic field of galactic clusters and on the possible MCP velocity distribution result in a tight model-independent bound $\milli < 10^{-14} (\Mmilli/$GeV) \cite{Kadota:2016tqq}.

In contrast, the existing laboratory experiments dedicated to MCP searches result in bounds that strongly depend on the MCP masses by exploiting different MCP production mechanisms. 
For example, the strongest limits for MCP masses below the MeV scale
come from the study of orthopositronium decays into invisible states,
or from the 
comparison of the Lamb-shift measurements with QED predictions, which sets the limit  $\milli < {\cal O}(10^{-4})$.
Direct laboratory bounds on MCP couplings and masses have also been cast by accelerator experiments \cite{Davidson:1991si} through the ``beam dump'' technique \cite{Bjorken:2009mm}, yielding $\milli  < 3\times 10^{-4}$ for MCP masses up to 100 MeV.

Experiments studying the propagation of polarized light in a strong 
magnetic field also constrain the MCP pair production. In fact, provided that the photon energy beams (in the eV range) satisfy the condition $\omega > 2 \Mmilli$,  MCP induce an observable ellipticity $\psi$ of the outgoing beam through vacuum magnetic birefringence and dichroism \cite{Tsai:1975iz,Gies:2006ca}.
The measured upper bounds on $\psi$ from the BFRT experiment \cite{Cameron:1993mr} first and PVLAS \cite{Zavattini:2005tm} later, were then used to set an upper limit on millicharge $\milli < {\cal O}(10^{-6})$ for MCP masses below 
$\Mmilli< {\cal O}(10^{-1}\mathrm{eV})$.
More recently, new experimental proposals aim to investigate the Schwinger pair production of MCP in the strong electric field of cavities used in particle accelerators \cite{Lilje:2004ib}. These can potentially improve 
the upper bound up to $\milli < 10^{-6}$ using present cavities at TESLA \cite{Lilje:2004ib} and up to $\milli < {\cal O}(10^{-7})$ with near-future cavities, for MCP masses below $10^{-3}$ eV \cite{Gies:2006hv}.

However, no dedicated experiments targeting the direct MCP pair production in inelastic photon-photon scatterings have been proposed to date.
Indeed, being stable, pair-produced MCP would escape the detector without interacting, leaving a signature only in missing energy. Nonetheless,  dedicated experiments based on interferometric techniques with polarized photon beams have the potential to reveal the direct MCP production for $\Mmilli < {\cal O}({\rm eV})$. 

In this framework, the aim of the paper is to perform a complete study of the production of MCP in collisions of polarized photon beams. In particular, we identify suitable polarization observables to disentangle the production of millicharged scalars, fermions and vector bosons. These results can be used in many applications ranging from the investigations of sub-eV MCP in laser experiments to the search of heavy MCP particles above the GeV scale at the future polarized gamma-gamma collider facilities \cite{Ginzburg:1981vm,Ginzburg:1982yr,Telnov:2013bpa}.

The paper is organized as follows. In Sec. II we present the analytical
results for the photon polarized cross sections into a pair of MCP of spin 0, 1/2 and 1, identifying dedicated observables as well.
In Sec. III we discuss the phenomenological implications of these results, while our conclusions are reported in Sec. IV. The Appendix gives details regarding the interferometric detection method sketched in the paper and provides a first estimate of the signal-to-noise ratio that can be obtained with this technique.
\vspace{-0.5cm}

\section{Polarization-dependent cross sections}
\label{sec:cross_sections}

We now analyze the amplitudes and cross sections for polarized photon-photon scatterings into a pair of MCP $\mu_{\s}\bar{\mu}_{\s}$ of spin $S$,
\bea
\gamma(p_1,h_1)\, \gamma(p_1,h_1) &\to& \mu_{\s}(p_3) \bar{\mu}_{\s}(p_4)\, ,
\label{process}
\eea
where $p_i$ with $i=1-4$ are the corresponding 4-momenta and $h_{1,2}$ indicate
the helicities of initial photons.

The generic amplitude for a process with two initial-state photons can be written as
\begin{align}
	\mathcal{M}_{h_1h_2} &= \epsilon_{h_{1}}^{\alpha}(p_1)\epsilon_{h_{2}}^{\beta}(p_2)\mathcal{M}_{\alpha\beta},
\label{ampl}
\end{align}
where $\epsilon_{h_{1,2}}^{\alpha}(p_{1,2})$ are the polarization vectors of the initial photons with helicities $h_{1,2} = \pm1$. For the production of two MCP with mass $\Mmilli$, the polarized differential cross section can be decomposed as
\begin{align}
	\td \sigma_{h_1h_2} 
&	=	 \frac{\td t}{16 \pi s^{2}}|\mathcal{M}_{h_1h_2}|^{2}
	\nonumber\\
&	\equiv	\td (\sigma + h_1 \, \sigma_{A1} + h_2 \, \sigma_{A2} + h_1 h_2 \, \sigma_{AA}).
\label{dsigma}
\end{align}
The angular dependence in the center-of-mass (c.m.) frame follows from $|\td t| = \beta s/2 \, \td(\cos(\theta))$, where $\beta = \sqrt{1 - 4 \Mmilli^2/s}$ is the speed of the final particles. The quantities on the right-hand side of Eq.~(\ref{dsigma}) correspond to the differential cross section ($d\sigma$) and the differential asymmetries ($d\sigma_A$ and $d\sigma_{AA}$), whose explicit form is obtained by inverting Eq.~\eqref{dsigma}:
\begin{subequations}
\begin{align}
	\td \sigma		&= \frac{1}{4}\td(\sigma_{++} + \sigma_{--} + \sigma_{+-} + \sigma_{-+}),
	\\
	\td \sigma_{A1} &= \frac{1}{4}\td(\sigma_{++} - \sigma_{--} + \sigma_{+-} - \sigma_{-+}),
	\\
	\td \sigma_{A2} &= \frac{1}{4}\td(\sigma_{++} - \sigma_{--} - \sigma_{+-} + \sigma_{-+}),
	\\
	\td \sigma_{AA} &= \frac{1}{4}\td(\sigma_{++} + \sigma_{--} - \sigma_{+-} - \sigma_{-+}).
\end{align}
\end{subequations}
Parity conservation implies $\td \sigma_{A1} = 0 = \td \sigma_{A2}$ \footnote{Algebraically, this follows from the fact that without parity violation the amplitude is a linear combination of the metric and tensor products of the momenta. Since there are only three independent momenta in this process, all the contractions with a single Levi-Civita tensor (contained in $A_{p_1,p_2}^{\alpha\alpha'}$) therefore vanish. The situation is different if  two or more Levi-Civita tensors are involved, as they can be contracted with each other.} thus, if we restrict ourselves to the case of parity-conserving interactions, the generic polarized total cross section can be expressed as follows
\bea
	\sigma_{h_1h_2} &=& \sigma (1 + \mathcal{A} \, h_1 h_2). 
\label{def_polasym}
\eea
In the above equation we defined the polarization asymmetry as
\begin{align}
	\mathcal{A} 
	\equiv \frac{\sigma_{AA}}{\sigma}
	= -\frac{\int \td t\, \mathcal{M}_{\alpha\beta}\mathcal{M}^{*}_{\alpha'\beta'} A_{p_1,p_2}^{\alpha\alpha'}			A_{p_1,p_2}^{\beta\beta'}}{\int \td t\, \mathcal{M}_{\alpha\beta}\mathcal{M}^{*\alpha\beta}}\, ,
\label{ASYM}
\end{align}
with the antisymmetric tensor
\begin{align}
	A^{\mu\nu}_{p_1,p_2} = -A^{\mu\nu}_{p_2,p_1} 
&	\,=\,  \frac{i\epsilon^{\mu\nu\rho\sigma}p_{1\rho}p_{2\sigma}}{p_{1}p_{2}}.
\end{align}
Here, $\epsilon^{\mu\nu\rho\sigma}$ denotes the totally antisymmetric tensor; we take $\epsilon^{01234}=1$. Notice that $A^{\mu\nu}_{p_1,p_2}$ can be equivalently defined as the difference between the density matrices of left and right polarization states, $A^{\mu\nu} = \epsilon_{+}^{\mu}\epsilon_{+}^{*\nu} - \epsilon_{-}^{\mu}\epsilon_{-}^{*\nu}$, and is thus identified as the observable to track the left-right polarization asymmetry of a given particle species.

\begin{figure}[h]
	\centering
		\includegraphics[trim={5cm 14cm 0 5cm }, clip, width=.65\textwidth]{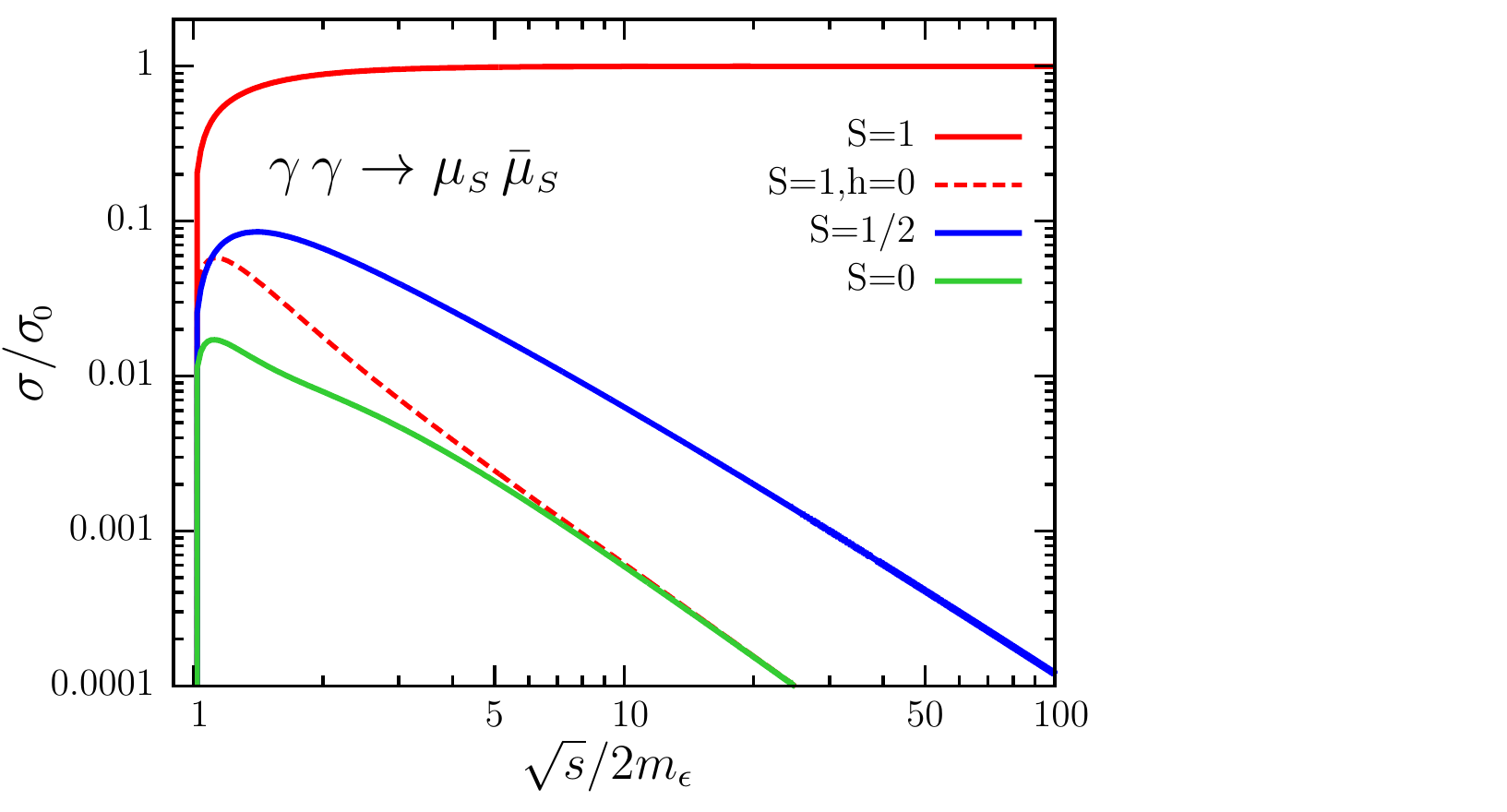}
		\includegraphics[trim={5cm 14cm 0 5cm }, clip, width=.65\textwidth]{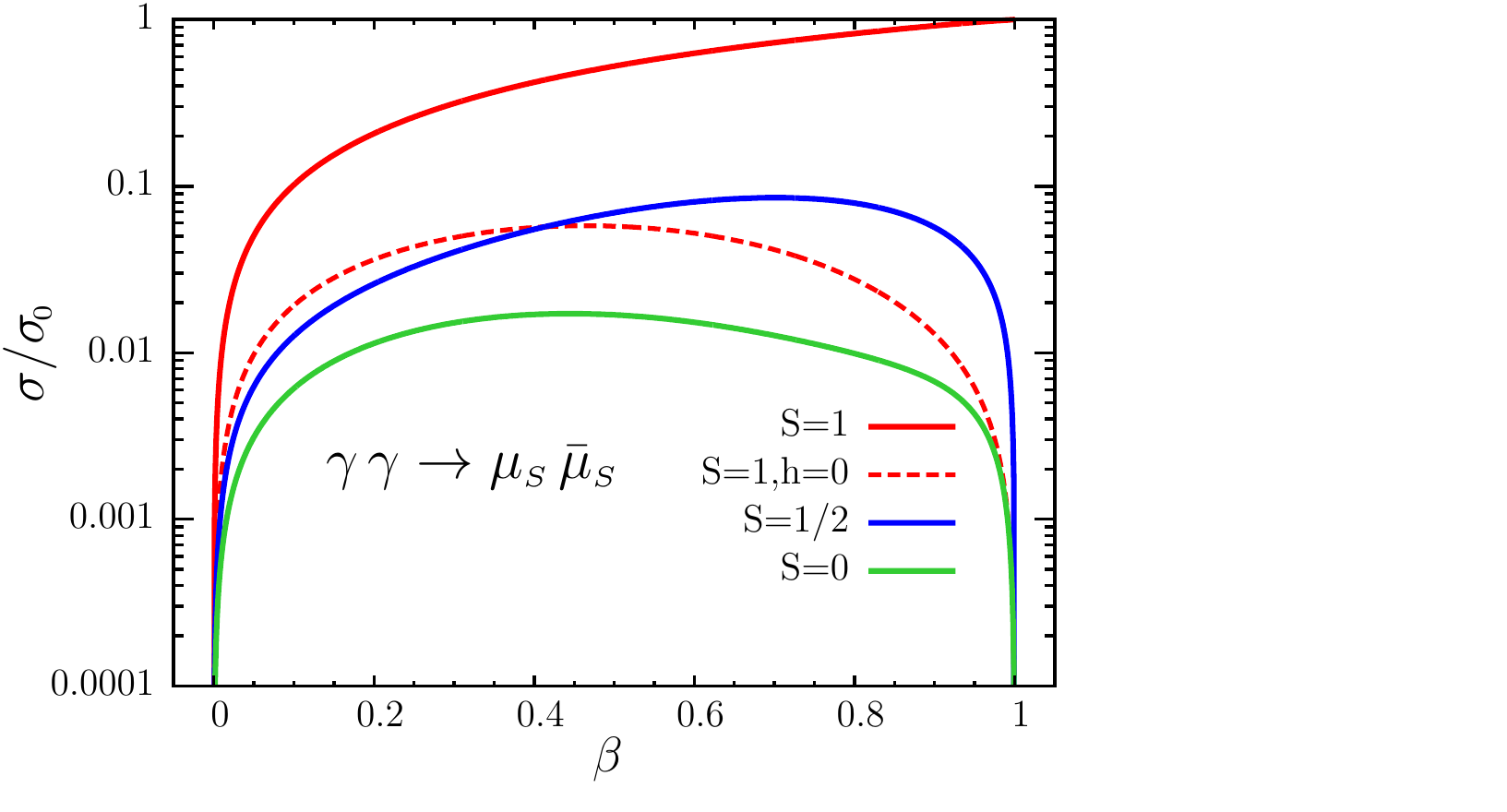}
\vskip-0.5cm
	\caption{Total unpolarized cross sections for $\gamma\gamma\to \mu_{\s}\bar{\mu}_{\s}$, normalized to $\sigma_{0}$ [Eq.(\ref{limcross})], for MCP $\mu_{\s}$ of spin $S = 0$ (green), $S=1/2$ (blue), and $S = 1$ (red) in continuous lines versus $\sqrt{s}/2\Mmilli$ and $\beta$, respectively, in the 
upper and lower panels. The dashed red line corresponds instead to 
the longitudinal polarization for $S=1$ MCP.
The analytic expressions for the cross sections are presented in Table~\eqref{tab:tot_cs}.}
	\label{fig:sigma_tot}
\end{figure}

Some remarks on the polarized cross sections for MCP production at the mass threshold follow. 
Being massless, the initial photons carry spin $\pm 1$, while the spin-0 state is forbidden; consequently, the initial two-photon state has either spin 0 or spin 2. In the 
c.m. frame, where the photon momenta are back-to-back, the spin-0 configuration corresponds to helicities\footnote{Notice that matching helicities in the c.m. frame correspond to anti-aligned spins as the photons propagate in opposite directions.} $++$ and $--$, while the spin-2 configuration is given by the helicities $+-$ and $-+$. If parity is conserved, it follows that $\sigma_{++} = \sigma_{--}$ and $\sigma_{+-} = \sigma_{-+}$.
If the final-state particles are produced at the threshold, almost at rest, the orbital angular momentum is negligible, and the spin alignment of the produced particles is therefore predictable. 
For production at the threshold, the spin-2 configuration is clearly forbidden for spin-0 and spin-1/2 MCP ($\sigma_{+-} = \sigma_{-+} = 0$) by the conservation of angular momentum. Hence, it follows that
\begin{align}
	\left.\mathcal{A}_{S=0,1/2}\right|_{\beta \to 0} = 1 \,;
\label{eq:lowS_th}
\end{align}
however, this argument clearly cannot be applied to spin-1 MCP final states, which clearly contain a valid spin-2 configuration.

\begin{table*}[ht]
\centering
\begin{tabular}{|p{0.08\linewidth}|p{0.45\linewidth}|p{0.45\linewidth}|}
\hline
$S$	
& $(\td \sigma / \td y)/\sigma_{0}$ 
& $(\td \sigma_{AA} / \td y)/\sigma_{0}$	\\
\hline
&&\\
$0 $	
& $\dfrac{\beta(1-\beta^2)}{32(1 - y^2 \beta^2)^2}((1 - \beta^2)^{2} + (1 - y^2)^{2} \beta^4)$ 
& $\dfrac{\beta(1-\beta^2)}{32(1 - y^2 \beta^2)^2}(1 - y^2 \beta^2)(1 - (2 - y^2) \beta^2)$ \\
&&\\
$1/2$	
& $\dfrac{\beta(1-\beta^2)}{16(1 - y^2 \beta^2)^2}(1 - \beta^4 + 2(1 - y^2)^{2} \beta^2 - (1 - y^2)^{2} \beta^4)$ 
& $\dfrac{\beta(1-\beta^2)}{16(1 - y^2 \beta^2)^2}(1 + y^2 \beta^2) (1 - (2 - y^2) \beta^2)$	\\
&&\\
$1$		
& $\dfrac{\beta(1-\beta^2)}{32(1 - y^2 \beta^2)^2}(19 + 2 (-3 + 8 y^2) \beta^2 + 3 (2 - 2 y^2 + y^4) \beta^4)$ 
& $-\dfrac{\beta(1-\beta^2)}{32(1 - y^2 \beta^2)^2}(13 + 3 y^2 \beta^2) (1 - (2 - y^2) \beta^2)$ \\
&&\\
\hline
\end{tabular}
\caption{Differential cross sections $\td \sigma / \td \cos(\theta)$ and differential polarization asymmetry $\td \sigma_{AA} / \td \cos(\theta)$ for photon-photon scattering into spin $S = 0,1/2,1$ MCP. The overall dimensional  normalization factor is $\sigma_{0} = 8\pi \alpha^{2} \milli^4/\Mmilli^2$ and $y = \cos(\theta)$ is the cosine of the scattering angle in the c.m. frame. }
\label{tab:diff_cs}
\end{table*}

The high energy asymptotic behavior, $s \gg \Mmilli$, of the total cross section of vector MCP is also qualitatively different from the scalar and fermion MCP ones. By using dimensional analysis, it could be expected that in this limit the total cross sections scale as $\sigma \sim 1/s$ for all spins.
However, while this assumption holds for the standard $\sigma_{S=0}$ and  $\sigma_{S=1/2}$ cases, it breaks down in a unitary theory of spin-1 MCP, where a remarkable phenomenon appears: The asymptotic value of the total cross section $\sigma_{S=1}$ tends to a constant given by
\begin{align}\label{limcross} 
	\lim_{s/\Mmilli\to \infty} \sigma_{S=1} 
 	=\,  \frac{8 \pi \alpha^2 \milli^4}{m_\epsilon^2}\,
 	\equiv \sigma_{0}.
\end{align}
This result arises from a collinear singularity in the differential cross section 
\bea
\frac{d\sigma}{dt} &\sim& \frac{1}{(t-\Mmilli^2)^2}
\label{diff}
\eea 
which is peaked for $t \sim 0$. The total cross section is, however, finite because the integration over $t$ is regularized by the MCP mass $\Mmilli$. 
The nonvanishing term in the right-hand side of Eq.~(\ref{diff}) appears because of the presence of the $\gamma V^+V^-$ vertex for an on-shell photon, whereas in the case of scalars and fermions, the double pole of Eq.~(\ref{diff}) vanishes because of the EM Ward identities (WI).
Interestingly, in the case of vector MCP, the WI do not ensure the cancellation of this term, which is present only if the tree-level gyromagnetic factor for the spin-1 theory is equal to $g=2$. 
The value of the gyromagnetic factor is set by requiring the unitarity of the theory at tree level \cite{Ferrara:1992yc}, implying that all the consistent theories of spin-1 particles interacting with the EM field, such as the one in \cite{Gabrielli:2015hua}, necessarily yield this phenomenon.
A tangible example of this characteristic asymptotic behavior is provided by the $W$ pair production through photon-photon scatterings in the SM, $\gamma \gamma \to W^+W^-$, where $g^{W}=2$ and $\sigma_{s\to\infty}(WW)=\frac{8\pi\alpha^2}{m_W^2}= 80.8~ \rm pb$, with $m_W$ the $W^{\pm}$ mass, while for $\alpha$ 
the low energy value should be used \cite{Katuya:1982ga,Denner:1995jv}. 
We stress that this is a general behavior of interacting spin-1 gauge theories based on non-Abelian gauge groups \cite{Gabrielli:2015hua}, which is also manifest in the pure gluon-gluon scatterings of QCD.\footnote{In this case, since the gluon is massless, the collinear singularity of Eq.~(\ref{diff}), affecting the total $gg\to gg$ cross section, is understood to be regularized by the QCD confinement scale $\Lambda_{QCD}$.}

We remark that higher order corrections are expected to induce a mild energy dependence in the asymptotic energy limit of the cross section, for instance after the Sudakov resummation of the large log terms 
$(\varepsilon^2\alpha\log{(s/m^2)})^n$ is included in the computation. However, these contributions are still unknown for the process we consider, and a dedicated study is thus needed to assess their relevance in this framework, a nontrivial task which goes beyond the purposes of the present paper.

\begin{table*}[ht]
\centering
\begin{tabular}{|p{0.08\linewidth}|p{0.55\linewidth}|p{0.15\linewidth}|p{0.15\linewidth}|}
\hline
$S$	
& $\sigma/\sigma_{0}$ 
& $\sigma/\sigma_{0} |_{\beta \to 0}$
& $\sigma/\sigma_{0} |_{s \to \infty}$	\\
\hline
&&&\\
$0 $	
&  $\dfrac{1}{16}\beta (1-\beta^2) \left( 2 - \beta^2 - (1-\beta^4) \beta^{-1}\atanh(\beta) \right)$
&  $\dfrac{1}{16}\beta$
&  $\dfrac{\Mmilli^{2}}{4s}$\\
&&&\\
$1/2$	
&  $\dfrac{1}{8}\beta (1-\beta ^2) \left( -2 + \beta^2 + (3-\beta^4) \beta^{-1}\atanh(\beta) \right)$
&  $\dfrac{1}{8}\beta$
&  $\dfrac{\Mmilli^{2}}{2s}\ln\left(\dfrac{s}{\Mmilli^{2}}\right)$\\
&&&\\
$1$		
&  $\dfrac{3}{16}\beta (1-\beta ^2) \left(2-\beta^2 - (1-\beta^4) \beta^{-1}\atanh(\beta)\right) + \beta$,
&  $\dfrac{19}{16}\beta$
&  $1$\\
&&&\\
$1\, (h_{\milli} = 0)$		
&  $\dfrac{1}{16}\beta(1-\beta ^2)\bigg(6\beta^{-4} - 8 \beta^{-2} + 4 - \beta^{2} +$ \newline
   \hspace{2cm}$+ \left(-6\beta^{-4} + 10\beta^{-2} - 3 - 2\beta^{2} + \beta^{4}\right) \beta^{-1}\atanh(\beta)\bigg)$
&  $\dfrac{47}{240}\beta$
&  $\dfrac{\Mmilli^{2}}{4s}$\\
&&&\\
\hline
\end{tabular}
\caption{Total cross sections $\sigma$ and corresponding asymptotic limits for photon-photon scattering into spin $S = 0,1/2,1$ MCP. The dimensional overall factor is $\sigma_{0} = 8\pi \alpha^{2} \milli^4/\Mmilli^2$. These cross sections are plotted in Fig.~\ref{fig:sigma_tot}. }
\label{tab:tot_cs}
\end{table*}

That being said, the analytical results for the polarized differential cross sections in the c.m. frame, for the spin-0, 1/2 and 1 MCP cases, are reported in Table \ref{tab:diff_cs}. The corresponding total cross sections are obtained by integrating over $\cos{\theta} \equiv y$. The total unpolarized cross sections and their asymptotic limits for the production of MCP with spin $S = 0,1/2,1$, as well as for the longitudinal polarization of $S = 1$ MCP, are reported in Table \eqref{tab:tot_cs} and plotted in Fig.~\ref{fig:sigma_tot}. We remark that the cross section of vector MCP $\sigma_{S=1}$ differs from the corresponding scalar cross section $\sigma_{S=0}$ by a factor of 3, on top of an additional term that tends to a constant in the high energy limit:
\begin{align}
	\sigma_{S=1} = 3 \, \sigma_{S=0} + \beta \sigma_{0}\, .
\end{align}	
Notice also that, in the same limit, the scalar MCP cross section matches 
the one for the longitudinal polarization of the vector MCP case, as expected from the Goldstone boson equivalence theorem.

\begin{figure}[h]
	\centering
	\includegraphics[trim={5cm 14cm 0 5cm }, clip, width=.65\textwidth]{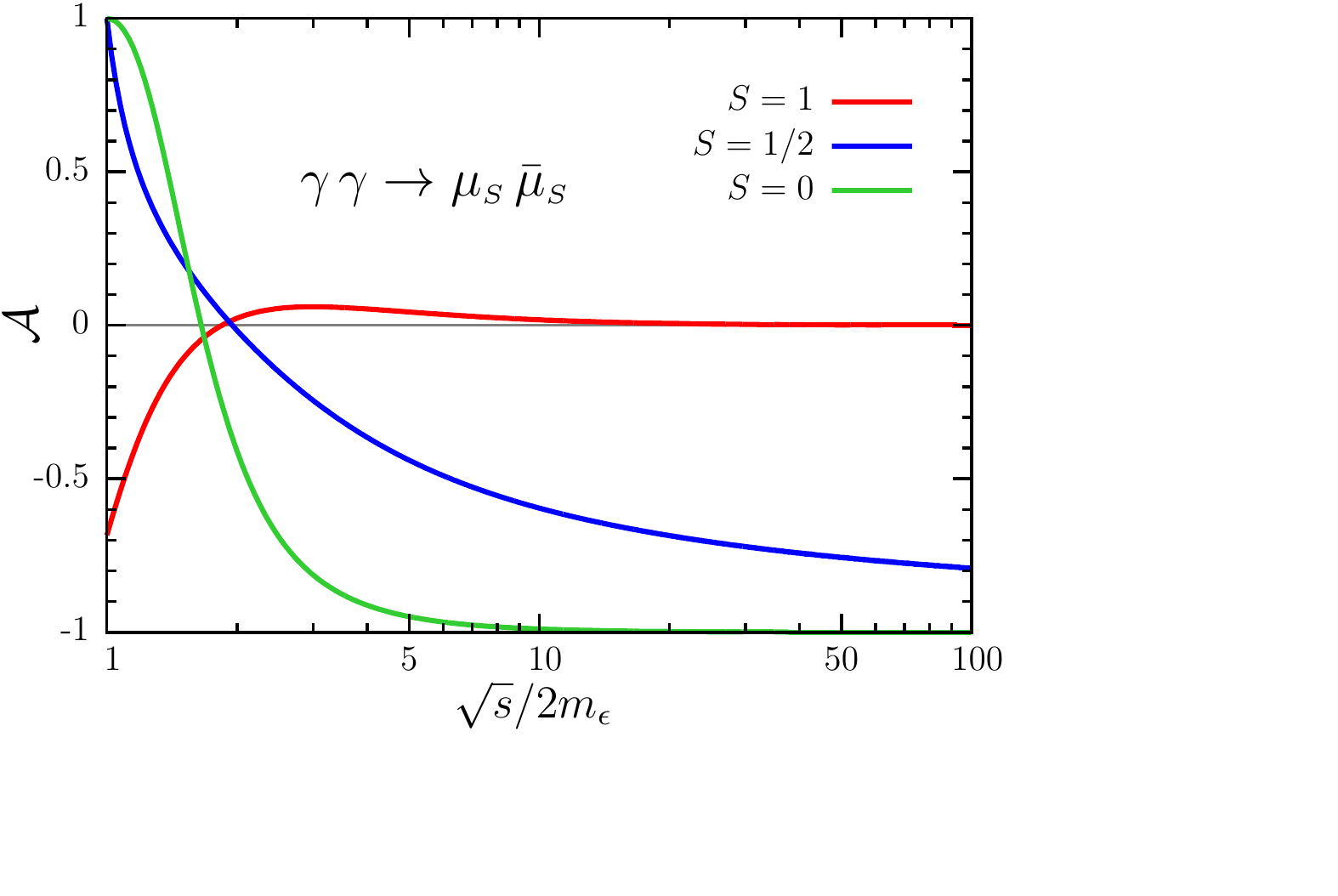}
\includegraphics[trim={5cm 14cm 0 5cm }, clip, width=.65\textwidth]{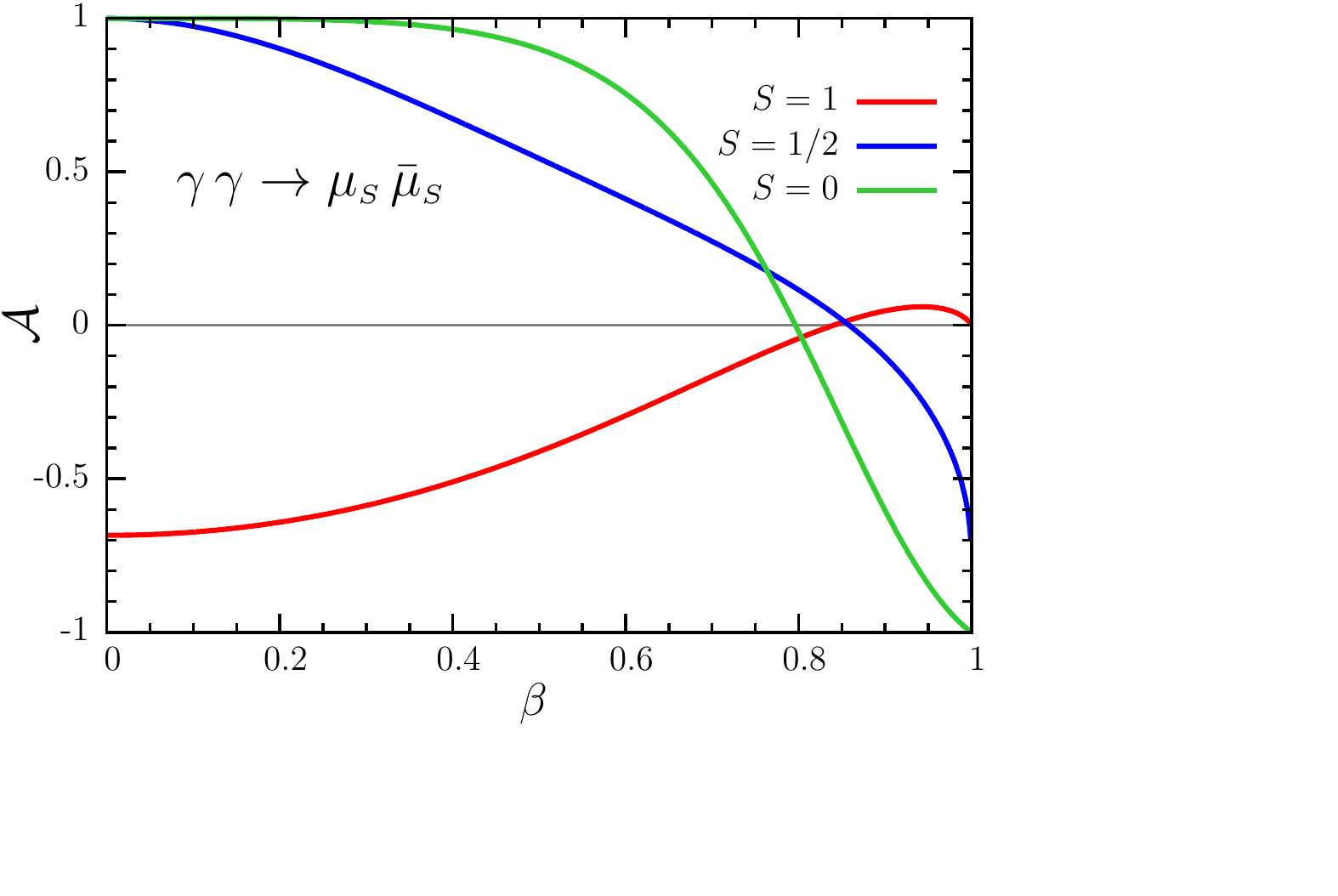}
\vskip-0.5cm
	\caption{Polarization asymmetry $\mathcal{A}$ (\ref{ASYM}) for $S = 0$ (green), $S=1/2$ (blue), and $S = 1$ (red), versus $\sqrt{s}/2\Mmilli$ and $\beta$ in the upper and lower panels, respectively. The analytic expressions of the asymmetries are presented in Table~\eqref{tab:tot_A}.}
	\label{fig:sigma_asym}
\end{figure}

The analytical results for the polarization asymmetry $\mathcal{A}$,  Eqs.~\eqref{def_polasym} and \eqref{ASYM}, and its asymptotic behavior are given in Table~\eqref{tab:tot_A} and shown in Fig.~\ref{fig:sigma_asym}. As we can see, the direct calculation confirms the threshold values, $\beta \to 0$, that we derived from general considerations on angular-momentum conservation Eq.~\eqref{eq:lowS_th}. 

\begin{table*}[ht]
\centering
\begin{tabular}{|p{0.08\linewidth}|p{0.55\linewidth}|p{0.15\linewidth}|p{0.15\linewidth}|}
\hline
$S$	
& $\mathcal{A}$
& $\mathcal{A} |_{\beta \to 0}$
& $\mathcal{A} |_{s \to \infty}$	\\
\hline
&&&\\
$0$	
&  $\dfrac{1 - 2(1-\beta^2)\beta^{-1}\atanh(\beta)}{-2 + \beta^2 + (1 -\beta^4)\beta^{-1}\atanh(\beta)}$
&  $1$
&  $-1$\\
&&&\\
$1/2$	
&  $\dfrac{3 - 2\beta^{-1}\atanh(\beta)}{- 2 + \beta^2 + (3 - \beta^4) \beta^{-1}\atanh(\beta)}$
&  $1$
&  $-1$\\
&&&\\
$1$		
&  $\dfrac{(1-\beta ^2) (-19 + 2(3 + 5\beta^2)\beta^{-1}\atanh(\beta))}{22 - 9\beta^2 + 3\beta^4 - 3(1 - \beta^2)(1-\beta^4) \beta^{-1}\atanh(\beta)}$
&  $-\dfrac{13}{19}$
&  $\dfrac{2\Mmilli^{2}}{s} \ln\left(\frac{s}{\Mmilli^{2}}\right)$\\
&&&\\
\hline
\end{tabular}
\caption{Polarization asymmetries $\mathcal{A}$ of Fig.~\eqref{fig:sigma_asym} and their asymptotic limits for photon-photon scattering into spin $S = 0,1/2,1$ MCP.}
\label{tab:tot_A}
\end{table*}

Figure~\ref{fig:sigma_tot} shows the unpolarized total cross sections, normalized to the asymptotic value of the total cross section for the spin-1 case, $\sigma_{0}$, versus $\sqrt{s}/(2\Mmilli)$ (upper panel) and  $\beta$ (lower panel), for the MCP spin cases $S=0,1/2,1$. For comparison, we also show the behavior of the spin $S=1$ longitudinal component ($h=0)$. 
As we can see from these results, for the same values of energy, MCP mass $\Mmilli$, and electric charge $\milli$, there is a well-defined hierarchy in the cross sections of $S=0,1/2,1$. 
In particular, assuming the same energy $\sqrt{s}$, mass, and couplings, we have
\bea
\sigma_{S=1} > \sigma_{S=1/2} > \sigma_{S=0}\, ,
\label{hierarchy}
\eea
for all $\sqrt{s}\ge 2\Mmilli$. 
Moreover,  for $s>4m_{\milli}$  the $\sigma_{S=1}$ largely dominates over the other cross sections $\sigma_{S=0,1/2}$ which decrease as $1/s$ already for $\sqrt{s} \gg m_{\milli}$. 
For comparative purposes, we also show the total cross section $\sigma_{S=1,h=0}$ related to the longitudinal component of the spin-1 MCP, which approaches the asymptotic high energy limit of the $\sigma_{S=0}$ curve as predicted by the Goldstone boson equivalence theorem.

The main message that can be drawn from these results is the following. 
By analyzing the inelastic photon-photon cross sections into invisible states, due to the constant asymptotic behavior of the $\sigma_{S=1}$ in the high energy limit, larger regions of the parameter space  $\milli, m_\epsilon$ could be probed for the spin $S=1$ scenario with respect to the cases of $S=0,1/2$ MCP, especially when small masses $\Mmilli$ are considered.
Following the results in Eq.~(\ref{hierarchy}), analogous conclusions hold for the $S=1/2$ case with respect to the spin $S=0$ one, although in this case both cross sections decrease as $1/s$ in the asymptotic limit $\sqrt{s}\gg \Mmilli$ and become almost insensitive to any mass $\Mmilli$ dependence for $\sqrt{s}\gsim 2\Mmilli$.

\begin{figure}[h]
\centering
\includegraphics[trim={5cm 14cm 0 5cm }, clip, width=.65\textwidth]{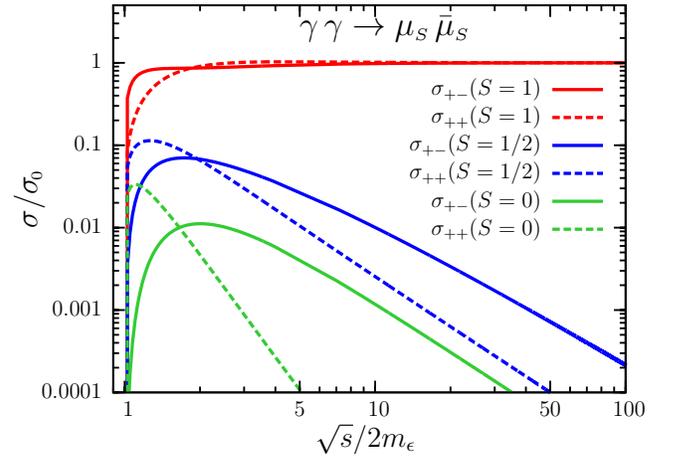}
\vskip-0.5cm
	\caption{Polarized cross sections for $S = 0$ (green), $S=1/2$ (blue), and $S = 1$ (red). A solid line denotes $\sigma_{+-} = \sigma_{-+}$ while the dashed line is $\sigma_{++} = \sigma_{--}$.}
	\label{fig:sigma_pol}
\end{figure}

Figure~\ref{fig:sigma_pol} shows the polarized cross sections $\sigma_{+-} = \sigma_{-+}$ and $\sigma_{++} = \sigma_{--}$, normalized to $\sigma_{0}$, as a function of $\sqrt{s}/\Mmilli$, for the various MCP spin $S=0,1/2,1$ configurations.
As we can see from these plots, the spin $S=0,1/2$ cases are characterized by an asymmetric behavior with respect to the initial photon polarization beams, with the opposite-helicity configurations $(h_1=+,h_2=-),\,(h_1=-,h_2=+)$ dominating over the same-helicity ones $(h_1=+,h_2=+)$, $(h_1=-,h_2=-)$ for $\sqrt{s}\gsim 4 \Mmilli$,  while the spin $S=1$ case is almost insensitive to the initial photon polarizations already for $\sqrt{s}\gsim 4 \Mmilli$.

In Fig. \ref{fig:sigma_asym} we show the polarization asymmetry $\mathcal{A}$ defined in Eq.~(\ref{ASYM}) for the $S=0,1/2,1$ cases, versus $\sqrt{s}/\Mmilli$ (upper panel) and $\beta$ (lower panel). 
These results show that the polarization asymmetry is very sensitive to the spin of MCP, and consequently, this quantity provides an efficient tool for disentangling the MCP spin once the corresponding signal has been detected.

\section{Phenomenological implications}
\label{sec:pheno}

\begin{figure}[h]
	\centering
\includegraphics[trim={4.5cm 14cm 0 5cm }, clip, width=.65\textwidth]{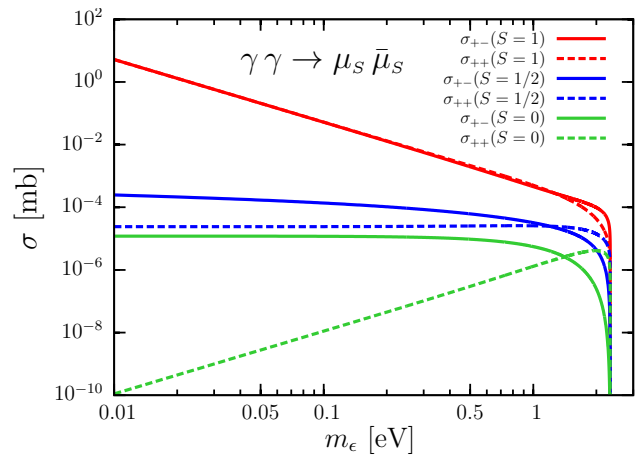}
		\vskip-0.5cm
	\caption{Mass dependence of the polarized cross sections for scattering of two 543 nm (2.33 eV) photons, with $\milli = 10^{-3}$. Solid and dashed lines denote $\sigma_{+-} = \sigma_{-+}$ and $\sigma_{++} = \sigma_{--}$, respectively, while the color code represents the spin as $S = 0$ (green), $S=1/2$ (blue), and $S = 1$ (red).}
	\label{fig:sigma_polm}
\end{figure}

As a first example of the possible applications of our results in dedicated experiments, we analyze the mass dependence of the polarized cross sections in the realistic case of two-photon scattering in the eV range. 
In particular, in Fig. \ref{fig:sigma_polm} we plot the total cross sections for the scattering of two 543 nm (2.33 eV) photons\footnote{The wavelength 543 nm has been chosen as an example because of the wide availability of 543 nm HeNe lasers. Any other common laser\textemdash
 such as the 532 nm frequency-doubled NdYAG laser \textemdash would lead to a very similar plot.}, for a typical millicharge value $\milli = 10^{-3}$, as a function of the MCP mass $\Mmilli <$ eV, for the spin cases $S=0,1/2,1$. 
These results show that the total polarized cross sections $\sigma_{++}=\sigma_{++}$ and $\sigma_{+-}=\sigma_{-+}$ for the spin $S=1$ MCP  are almost 
indistinguishable for  $\Mmilli<1.5$ eV, while they monotonically decrease by increasing the mass $\Mmilli$. 
A different behavior is observed for the $S=1/2$ case, where the cross section $\sigma_{++}$ dominates over $\sigma_{+-}$ for $\Mmilli\lsim 1.5$ eV and it is almost insensitive to the mass.
On the other hand, in the case of spin 0, we have that $\sigma_{++} > \sigma_{+-}$ for $\Mmilli\lsim 1.5$ eV, while the $\sigma_{++}$ remains almost constant, with $\sigma_{++} \simeq 5\times 10^{-5}$mb, for  $\Mmilli\lsim 1.5$ eV. 
These features can be easily understood by referring to the asymptotic energy limits of the polarized cross sections in Table \ref{tab:tot_cs}.

\begin{figure*}[!ht]
\begin{center}
\includegraphics[width=0.9\textwidth]{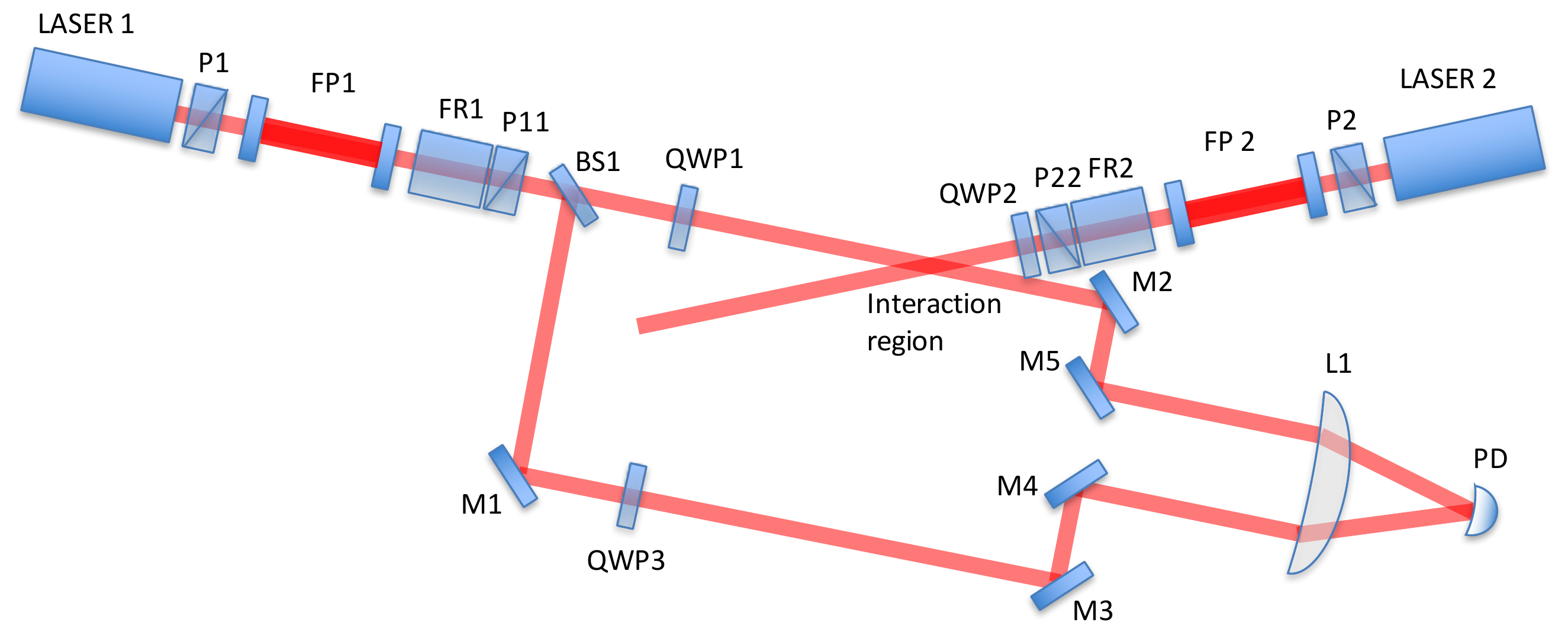}
\end{center}
\caption{A possible experimental layout.  Two laser beams (the main beam LASER 1 and the probe beam LASER 2) are linearly polarized by the polarizers P1 and P2. The Fabry-Perot cavities FP1 and FP2 are used to reduce the pointing noise of each laser and as components of a frequency-stabilizing scheme of the Pound-Drever-Hall type.  Notice that here frequency stabilization matters because of the interferometric detection scheme. The FR1 and FR2 Faraday cells and the polarizers P11 and P22 are used as amplitude modulators, while the quarter-wave plates QWP1--QWP3 produce circularly polarized beams. The polarizations are left- or right-handed according to the position of the slow and fast axes of the QWPs. The main beam from LASER 1 is split by a beam splitter BS1 into two beams that are balanced as much as possible (the optical elements used for balancing are omitted), where M1--M5 are optical mirrors. Then, the beams are combined by the convex lens L1, and the optical path lengths are tuned\textemdash for instance, by moving mirror M5 relative to M2\textemdash to have a dark fringe on the photodiode 
(PD). Intensity changes in one of the two balanced beams are then detected as deviations from this condition.}
\label{exp_setup}
\end{figure*}

\bigskip

But how could we best detect sub-eV MCP directly? We mentioned that experiments aimed at verifying the QED predictions on photon-photon scattering in the eV energy range, which use polarized light propagating in a strong magnetic field and Fabry-Perot cavities like the PVLAS experiment \cite{pvlas:2012}, can be used to set bounds on $S=0$ and $S=1/2$ MCP couplings and masses \cite{Gies:2006ca}. 
In fact, only the contributions to 
vacuum magnetic birefringence and dichroism effects of $S=0,1/2$ MCP have been analyzed in the literature so far, while the extension of these results to the $S=1$ case is in progress \cite{gmv}. 
Notice also that the measurement of vacuum magnetic dichroism induced by the pair production mechanism strongly depends on the external magnetic field and yet it does not provide a direct measurement of the cross section of photon-photon scatterings into MCP, nor does it provides information on the spin of the produced MCP.

So far, no experiment has tried to directly observe the inelastic photon-photon
scatterings into MCP. At low energy (i.e., in the optical energy range), this could, in principle, be achieved by directly measuring these scatterings with polarized laser beams and Fabry-Perot interferometers. 
Unfortunately, the latter are not well suited to this task because the antipropagating beams have anticorrelated polarizations, so a laser beam with right-handed circular polarization moving in one direction along the interferometer axis would collide with a reflected beam that is left-handed polarized. For this reason it is not possible to extract all of the polarized cross sections with this technique.

\begin{figure}[!ht]
\begin{center}
\includegraphics[width=.45\textwidth]{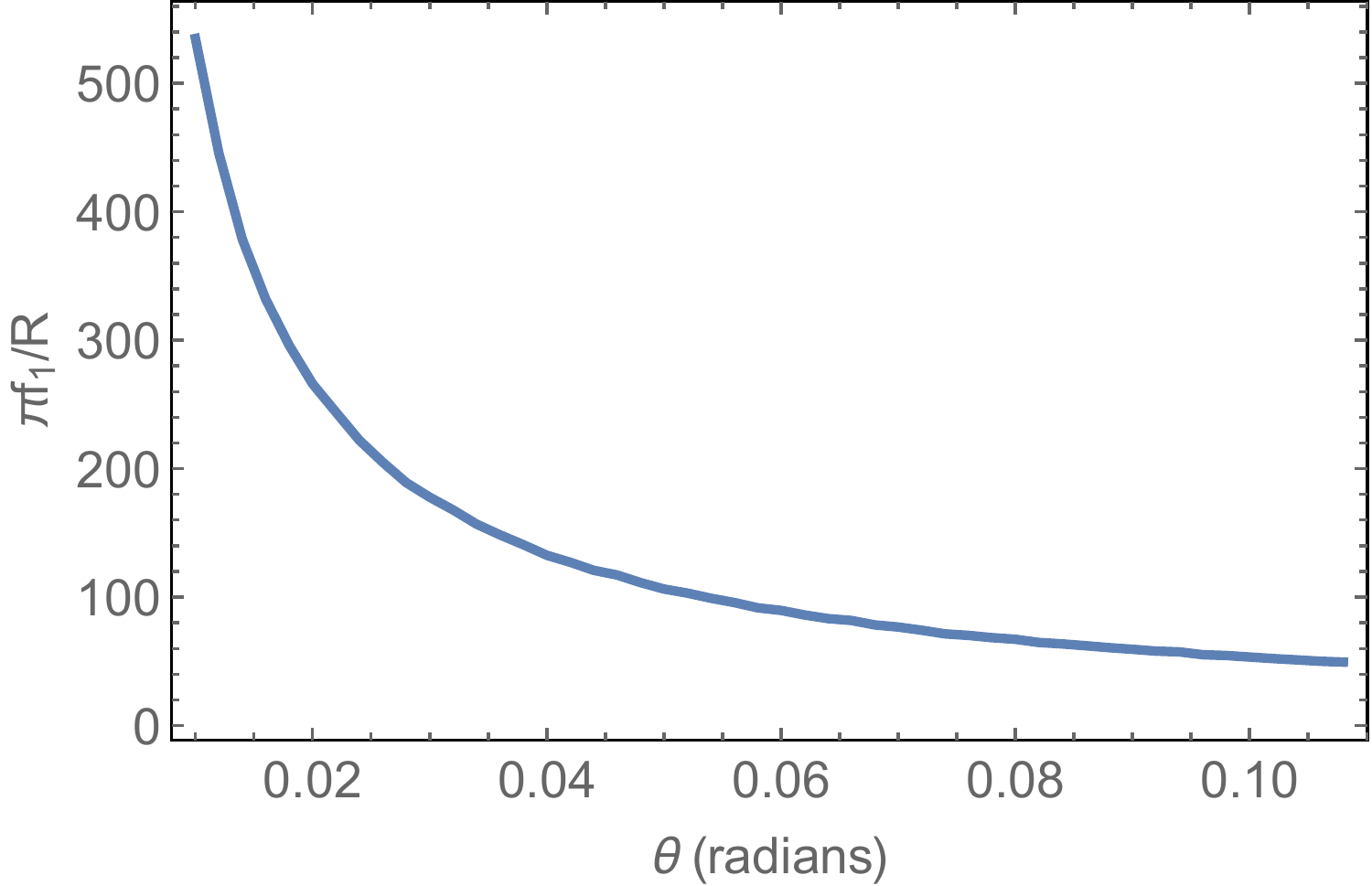}
\end{center}
\caption{Plot of the geometric factor in Eq. \eqref{intensity}, in normalized units vs the angle $\theta$ between the two beams. Here we assume that both the main beam 1 and the probe beam 2 are simple cylindrical beams with the same radius $R_1=R_2=R$.}
\label{GeomFact}
\end{figure}

An alternative method that gives full freedom in selecting the photon polarizations utilizes an arrangement with independent near-visible laser beams, as shown in the prototype experiment in Fig.\ref{exp_setup}. 
The layout is a very-low-energy ``photon-photon collider'' where two stabilized, polarized laser beams collide in a narrow region. In this arrangement a main laser beam scatters photons from a modulated probe beam: Any intensity change in the intensity of the main beam is detected with an interferometric scheme, as shown in Fig.\ref{exp_setup}, where we search for an  interaction by monitoring the intensity of the main beam at the modulation frequency of the probe beam. Possible intensity changes due to the beam-beam interaction are described by the equation 
\bea
I'_1(t) &=& I_1(t) - \frac{\sigma \lambda}{h c^2} f_1 I_1(t) I_2(t)
\label{intensity}
\eea
where $\sigma$ is one of the polarization-dependent cross sections defined above, $h$ is the Planck constant, $\lambda$ is the wavelength of each laser, $f_1$ is a geometric factor parametrizing the effective interaction region, which has the dimensions of a length and depends on the angle $\theta$ between the beams (see Fig. \ref{GeomFact}), and $I_{1,2}(t)$ are the time-dependent beam intensities. The time dependence in $I_{1,2}(t)$ originates from the modulation used to extract the weak signal out of the experimental noise, according to a time-tested scheme (see, e.g., Ref. \cite{pvlas:2012}). If only the probe beam is modulated 
\bea
I_{1}(t) &=& I^{(0)}_{1}  \\
I_{2}(t) &=& I^{(0)}_{2} \left[ 1+m_{2} \cos(\omega_{2} t + \varphi_{2}) \right]
\,,
\eea
with $\omega_{2}$ and $m_{2}$ the modulation frequency and index respectively,
we find, according to Eq.(\ref{intensity})
\bea
I_1'(t) &=& 
\left(I_1^{(0)}-\frac{\sigma \lambda}{h c^2} f_1  I_1^{(0)}I_2^{(0)}\right)  
\nonumber
\\
&-& \frac{\sigma \lambda}{h c^2} f_1  I_1^{(0)}I_2^{(0)}   m_2 \cos(\omega_2 t + \varphi_2) \, .
\label{mod}
\eea
We can see from Eq.(\ref{mod}) that, by extracting the modulation at frequency $\omega_2$ in the main beam intensity $I_1'(t)$, we can obtain the cross section $\sigma$. Further refinements with more complex modulation schemes also allow us to separate the signal from environmental noises and systematics. The basic interferometric detection scheme and its sensitivity are discussed in the Appendix.

\bigskip

To conclude this section, let us remark on the importance of these searches in relation to the physics beyond the SM. 
The phenomenology of MCP has a substantial reach, ranging from collider experiments to astroparticle physics and cosmology. 
The direct exploration of a possible light MCP sector, for instance, constitutes an independent check of the contemporary cosmological model. 
The usual thermal mechanisms invoked for the production of dark matter, in fact, normally yield a sizable abundance of light MCP once such a component is introduced in the theory. 
For their properties, sub-reMeV MCP generally alter the physics of big bang nucleosynthesis and the cosmic microwave background radiation, as well as the stellar evolution and possibly the DM halo evolution.  
Given that the dedicated observations put stringent indirect constraints on the parameter space of light MCP, the detection of these particles in laboratory experiments would have a radical impact on our current understanding of the Universe. 
In contrast, the potential discovery of a heavier MCP candidate at dedicated photon colliders would put forward a viable dark matter candidate with the potential to reconcile the Standard Model with the concordance model of cosmology. 
Heavy MCP dark matter candidates can also be investigated in direct dark matter searches, through the possible scattering of MCP on the SM particles mediated by photons, and in indirect searches, which aim to constrain the DM properties by analyzing the products of dark matter annihilations. 
Millicharged dark matter can also leave its imprints in the cosmic microwave background \cite{Dubovsky:2003yn}, in the large scale structure, or in collisions of galaxies and galaxy clusters where they may yield collisionless shocks once the dark matter streams through the astrophysical plasma \cite{Heikinheimo:2015kra}.
Furthermore, heavy DM MCP could be produced in collider experiments yielding the typical missing energy signature, as well as, for the spin-1 case, extra signals related to the required additional neutral gauge bosons. 
 
\section{Conclusions}
\label{sec:conclusions}
In the context of a new physics scenario that proposes new gauge and dark sectors, we have systematically analyzed the pair production of MCP in photon-photon scatterings $\gamma \gamma \to \mu_{\s}\bar{\mu}_{\s}$. MCP, which are
stable particles almost decoupled from ordinary matter, can be produced
in ordinary EM interactions owing to their small electric charge. In this framework, we have computed the corresponding differential and total cross sections for polarized initial photon beams in the case of MCP of spin $S=0,\, 1/2,\,1$.  Model-independent results were obtained for the corresponding cross sections by simply imposing the EM gauge invariance and unitarity of the theory.

Photon polarization asymmetries have also been analyzed for all the considered MCP spin 
cases: $S=0,1/2,1$. We found that these observables are very sensitive to the spin of the produced MCP and allow for its identification through measurements of photon-photon polarized cross sections.
All the results presented here have a general validity and can be applied to any range of MCP masses and photon energies, provided the kinematic conditions for the MCP production are satisfied.

In the case of $S=1$ MCP production, we show that, due to a collinear effect, the total cross section does not follow the canonical asymptotic behavior $\propto 1/s$ at high center-of-mass energy $\sqrt s$, in contrast to the $S=0,1/2$ cases. In particular, the total cross section tends to a constant proportional to the inverse of the MCP mass square in the limit $\sqrt s \gg m_{\epsilon}$, where $\sqrt s$ is the energy in the $\gamma \gamma$ center-of-mass frame. Therefore, ultralight vectorial $S=1$ MCP with masses $m_{\epsilon}\ll \sqrt s$ 
have potentially larger cross sections than MCP of spin $S=0,1/2$, given the same millicharge couplings and mass. This suggests that 
direct measurements of photon-photon interactions could prove a more sensitive tool for testing the production of vectorial MCP than the methods employed so far in the dedicated MCP searches.

To further investigate this possibility, we have considered a prototype 
experiment to directly measure the polarized $\gamma \gamma \to 
\mu_{\s}\bar{\mu}_{\s} $ cross sections. This is based on a suitable interferometric scheme with two stabilized polarized laser beams acting as low energy photon-photon collider. 

We conclude by remarking on the importance of these searches in connection with the physics beyond the Standard Model. Whereas measuring the properties of possible sub-MeV MCP would constitute an important indirect test of the contemporary cosmological model, multi-GeV MCP particles provide an interesting dark matter candidate that could be tested at future photon colliders and leave traces in direct and indirect detection experiments.

\section{Appendix}
Here we describe in further detail the interferometric detection method employed in the experimental setup of Fig. \ref{exp_setup} and estimate the sensitivity that can be reached by using state-of-the-art experimental equipment.

\subsection{Interferometric detection of photon-photon scattering and its sensitivity}

The interferential detection scheme operates around the dark fringe, as in gravitational wave interferometers (see, e.g., Ref.\cite{Maggiore}). This scheme has many advantages; 
for instance, it minimizes shot noise. 

Given that two sinusoidally varying electric fields with angular frequency $\omega$ and phase difference $\varphi$, once superposed, result in a total field
\bea
\mathbf{E}_\mathrm{tot}(\mathbf{x},t) &=& \mathbf{E}_1(\mathbf{x})\cos(\omega t) + \mathbf{E}_2(\mathbf{x})\cos(\omega t+\varphi), 
\eea
then the total irradiance is
\bea
I_\mathrm{tot} &=&\epsilon_0 c \langle |\mathbf{E}_\mathrm{tot}|^2 \rangle = I_1 + I_2 + 2 \sqrt{I_1 I_2} \cos\varphi
\eea
where $I_1 = \epsilon_0 c \langle |\mathbf{E}_{1}|^2  \rangle$ and $I_2 = \epsilon_0 c \langle |\mathbf{E}_{2}|^2  \rangle$ (see, e.g., Ref. \cite{Hecht}).

To operate the apparatus around the dark fringe, we let $\varphi = \pi$, so the interference is normally destructive, i.e., 
\bea
I_\mathrm{tot}  = I_1 + I_2 - 2 \sqrt{I_1 I_2}
\eea
Then, when the beams are balanced we get $I_1 = I_2 = I_0$,  so 
$I_\mathrm{tot}=0$. However, when the beams are off balance, namely 
$I_2=I_0-\delta I$ and $\delta I \ll I_{0}$, we have
\bea
I_\mathrm{tot} &=& 2I_0 - 2 \sqrt{I_0 (I_0-\delta I)} \approx \delta I\, .
\eea
Thus, we see that the irradiance measured in this scheme is exactly equal to the irradiance difference between the beams. A nonzero $\delta I$ can stem from the following: 
\begin{itemize}
\item[i)] {\it Actual physics.}\textemdash  This is the phenomenon we seek in the case in which $\delta I$ is due to an interaction with the probe beam because of the light-light scattering. This effect is modulated at the modulation frequency of the probe beam $\omega_0$.
\item[ii)] {\it Unbalance between the interferometer arms.}\textemdash  It may be that power is not exactly halved by the beam splitter BS1 in Fig. \ref{exp_setup}, or that there is a different beam attenuation because of mirrors and quarter-wave plates in the two arms of the interferometer. However, this is a dc term, which is eliminated by measuring the effect at the probe beam modulation frequency $\omega_0$.
\item[iii)] {\it Background noise.}\textemdash  There are several white noise sources that affect the accuracy of the measurement. These are an actual nuisance, and we list them below. 
\end{itemize}
The value of $\delta I$ in the MCP scenario discussed in this paper is evaluated in Eq. \eqref{mod}, and when we let  $I_1^{(0)}=I_0$ be the irradiance of the main beam and $I_2^{(0)}=I_\mathrm{probe}$ the irradiance of the probe beam, we find for the scattered irradiance
\bea
\label{dI}
\delta I &=& \frac{\sigma\lambda}{hc^2} f_1 I_0 I_\mathrm{probe} m_{\rm probe}\, .
\eea
where $m_2=m_{\rm probe}$ is the modulation index of the probe beam.
If  $R_{\rm det}$ is the spot size (radius) on the photodetector, then the total detected power $\delta P$ associated with the irradiance $\delta I$ is 
\bea
\delta P=\pi R^2_{\rm det} \delta I\, .
\eea
Moreover, the detector-amplifier pair converts power into a current $i=qP$, where $q$ is the quantum efficiency of the detector.
Finally, the signal-to-noise ratio of the experiment\textemdash evaluated as the mean square fluctuation of the signal divided by the mean-square fluctuation of noise\textemdash  is given by 
\bea
\mathrm{S/N}=\frac{\langle i_\mathrm{S}^2 \rangle}{i_\mathrm{RMS}^2} = \frac{(q \delta I \pi R_{\rm det}^2)^2}{i_\mathrm{RMS}^2}\, ,
\eea
where $R_{\rm det}$ is the radius of the spot size on the photodetector and $i_\mathrm{RMS}$ is the mean-square fluctuation of the photocurrent. Detection of the effect requires  that ${\rm S/N}>1$.

In the setup  of Fig. \ref{exp_setup},  the total mean-square fluctuation of the photocurrent is
\bea
i_\mathrm{RMS}^2 = i^2_{\rm shot}+i^2_{\rm dark}+i^2_{\rm therm}+i^2_{\rm rin}
\label{noise}
\eea
where all the individual terms depend on the frequency resolution $\Delta\nu = 1/T$, with $T$  the total data acquisition time; more specifically,
\bea
i_{\rm shot}^2 &=& 2 e (q \pi R_\mathrm{det}^2 \delta I) \Delta\nu 
\eea
is associated with the small shot noise due to the irradiance $\delta I$, where  $e$ is the elementary charge and where we assume that the interferometric scheme is well balanced so that there is no systematic contribution to the total $\delta I$. Note that
\bea
i_{\rm dark}^2  &=&\frac{V_\mathrm{dark}^2{\Delta\nu}}{G^2}\, 
\eea
 is related to the detector noise due to the dark current in the photodiode, where $V_\mathrm{dark}$ is the associated potential and $G$ is the transimpedance of the amplifier (as in \cite{Bregant}),
\bea
i_{\rm therm}^2&=&\frac{4k_B T \Delta\nu}{G}
\eea
 is the Johnson (thermal) noise  associated with the transimpedance of the amplifier, and finally
\bea
i_{rin}^2&=& q^2(\pi R_\mathrm{det}^2 I_0)^2 \mathrm{RIN}(\omega_0) \Delta\nu \, 
\eea
is related to the relative intensity noise (RIN)  due to the LASER [i.e., the fluctuations of the LASER output power, defined as $\mathrm{RIN}(\omega_0) = \langle \left(\Delta P(\omega_0)\right)^2 \rangle/P^2$, where $\langle \left(\Delta P(\omega_0)\right)^2 \rangle$ is the 
mean-square fluctuation of output power at frequency $\omega_0$ and $P$ is the average output power].

\subsection{Numerical estimates}
We now evaluate the magnitude of $\delta I$ and of the individual noise RMS values, on the basis of state-of-the-art parameter values.
 
To provide a first estimate, we assume that the intersection region roughly matches the position of the beam waist, and we take a beam waist $R= 1$ mm. Allowing for an angle between the beams that grants the positioning of all the equipment, such that $\pi f_1/R \approx 300$, we obtain $f_1 \approx 0.1$ m. 

Equation \eqref{dI} shows that the magnitude of the physical effect depends critically on the irradiance of the laser beams, and since we work with a modulation scheme, we must use modulated continuous wave (CW) lasers. In recent years fiber lasers have emerged as an extremely practical source of high-power laser radiation \cite{Zervas}, with average powers as high as tens of kW. Taking an average power of 50 kW both for the main and for the probe beam (as in recent industrial high-power fiber lasers produced by IPG Photonics, YLS series), a wavelength of $\lambda = 1075$ nm ($h\nu \approx 1.15$ eV) and the modulation index $m_\mathrm{probe} = 1$, we find the magnitude of the scattered irradiance 
\bea
\!\!\!\!\!\! \delta I  &\approx& \{1.8 \times 10^{-22}\; \mathrm{b}^{-1}(\mathrm{W/m^2})^{-1}\}  \sigma  I_0 I_\mathrm{probe} \, .
\eea
This scattered irradiance can also be expressed in terms of the number of scattered 
photons $N_{\gamma}$,
\bea
\delta I  &\approx& \{9.9\times 10^{-4}\; (N_{\gamma}/\mathrm{m}^{2}\mathrm{s})\;\mathrm{b}^{-1}(\mathrm{W/m^2})^{-2}\}   \sigma  I_0 I_\mathrm{probe}\, .
\nonumber
\eea 
The main beam and probe beam irradiance in the interaction region is  $I_0 = I_\mathrm{probe} \approx 1.6\times 10^{16}\;\mathrm{W/m^2}$, implying that 
\bea
 \delta I  &\approx& \{4.6 \times 10^{4}\; \mu\mathrm{b}^{-1}\}  \sigma   \times 
(\mathrm{W/m^2}).
\eea

Then, for the estimate of the noise terms, we take values in the published work of the PVLAS experiment \cite{Bregant}, so we have
\begin{itemize}
\item operating temperature: $T = 300$ K\, ,
\item quantum efficiency: $q = 0.7$ A/W\, ,
\item transimpedance (gain): $G = 10^7 $ V/A\, ,
\item photodiode noise: $V_\mathrm{dark} = 2\;\mu V/\sqrt\mathrm{Hz}$\, .
\end{itemize}
We take the RIN at low modulation frequency $\mathrm{RIN}(\omega_0)\approx 10^{-14}$ $1/\sqrt\mathrm{Hz}$ as in Ref. \cite{Spiegelberg}. 

Finally, assuming that the apparatus will be sufficiently stable over periods of the order of 10 days (i.e., considering a data acquisition time $\tau\approx 10^6$ s), we can obtain a frequency resolution $\Delta\nu=1/\tau \approx 1\;\mu$Hz. 

\begin{figure}[tbhp]
\centering
\includegraphics[width=8.7cm]{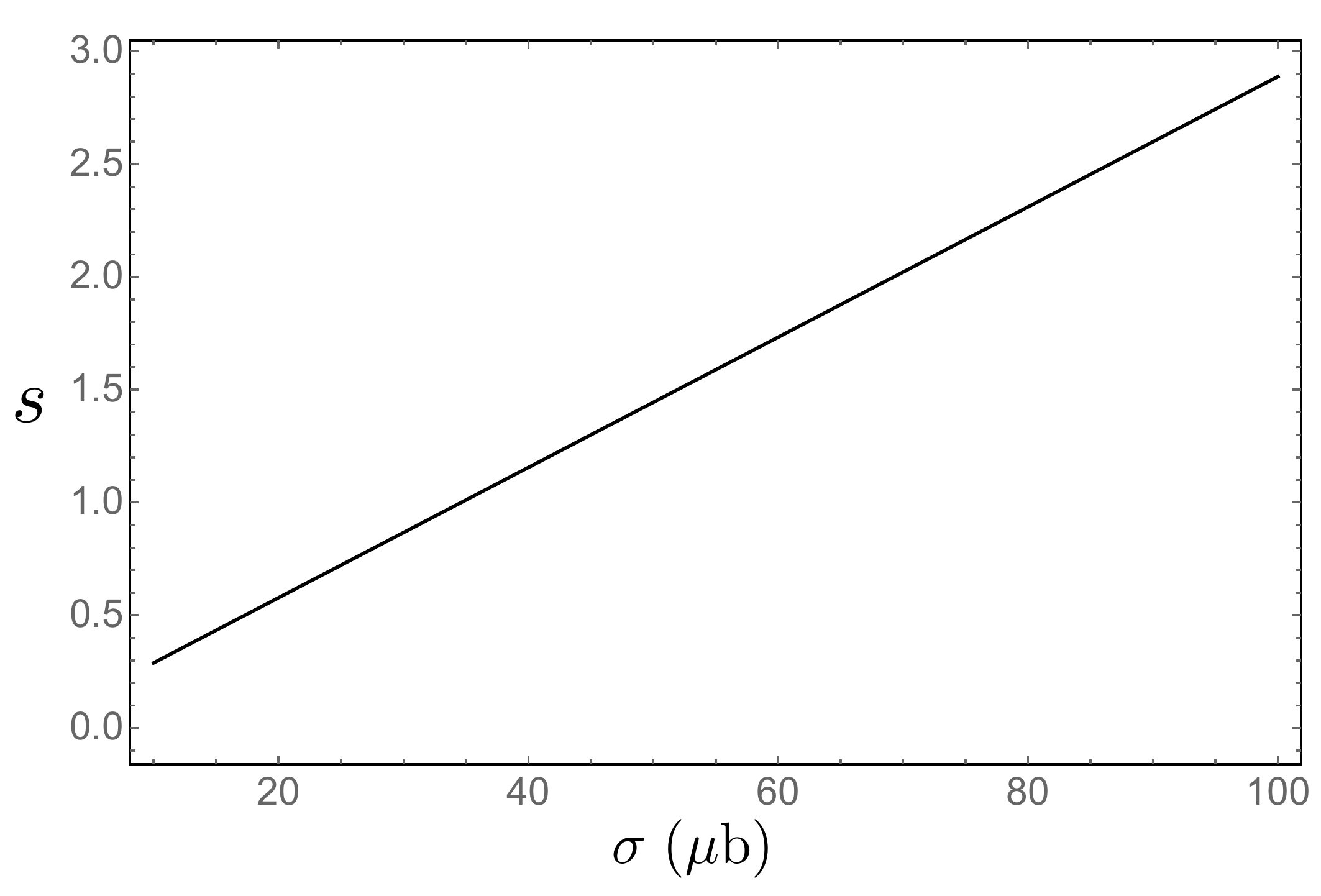}
\caption{Sensitivity $s=\sqrt{S/N}$ as a function of the cross section $\sigma$. This plot has been drawn with the parameters listed in the main text.}
\label{Signal2Noise}
\end{figure}

Adopting the above values, we find the dependence of the sensitivity $s=\sqrt{S/N}$ on the total cross section as shown in Fig. \ref{Signal2Noise}. From these results and the numbers quoted above, we obtain $s \approx 1$ at $\sigma \approx 35\;\mu$b. An increased sensitivity requires better data acquisition parameters\textemdash e.g., a lower RIN, which gives the largest contribution to noise\textemdash  higher intensities and longer data acquisition times. 

\smallskip

Eventually, this sensitivity can be translated into a sensitivity on the millicharge $\varepsilon$ depending on the values of the MCP masses. For instance,
according to the results in Fig.\ref{fig:sigma_polm} for spin-0 and 1/2 MCPs, a sensitivity of the order of $\mu$barn on the cross section would imply a sensitivity on the millicharge 
$\varepsilon \simeq {\cal O}( 10^{-3})$ for MCP masses below the eV scale. On the other hand, for the MCP spin-1 scenario, a much stronger sensitivity on $\varepsilon$ can be achieved at low MCP masses with respect to spin-0 and spin-1/2 cases, due to the constant asymptotic value of the cross section.

\section{Acknowledgments} 
E.G. would like to thank the CERN TH-Division and NICPB for their kind hospitality during the preparation of this work. L.M. acknowledges the Estonian Research Council for supporting his work through the grant No. PUTJD110. 

\end{document}